# Dynamics of ripple formation on silicon surfaces by ultrashort laser pulses in sub-ablation conditions

G. D. Tsibidis[1,*], M. Barberoglou[1], P. A. Loukakos[1], E. Stratakis[1,2], C. Fotakis[1,3]

[1]Institute of Electronic Structure and Laser (IESL), Foundation for Research and Technology (FORTH), N. Plastira 100, Vassilika Vouton, 70013, Heraklion, Crete, Greece

[2] Department of Materials Science and Technology, University of Crete, 710 03 Heraklion, Crete, Greece.

[3]Physics Department, University of Crete, Heraklion 71409, Crete, Greece

An investigation of ultrashort pulsed laser induced surface modification due to conditions that result in a superheated melted liquid layer and material evaporation are considered. To describe the surface modification occurring after cooling and resolidification of the melted layer and understand the underlying physical fundamental mechanisms, a unified model is presented to account for crater and subwavelength ripple formation based on a synergy of electron excitation and capillary waves solidification. The proposed theoretical framework aims to address the laser-material interaction in sub-ablation conditions and thus minimal mass removal in combination with a hydrodynamics-based scenario of the crater creation and ripple formation following surface irradiation with single and multiple pulses, respectively. The development of the periodic structures is attributed to the interference of the incident wave with a surface plasmon wave. Details of the surface morphology attained are elaborated as a function of the imposed conditions and results are tested against experimental data.

PACS numbers: 42.25.Hz, 64.70.D, 79.20.Ds, 64.70.fm

## I. INTRODUCTION

Silicon surface processing with ultra-short pulsed lasers has received considerable attention over the past decades due to its important technological applications, in particular in industry and medicine [1-11]. Rapid energy delivery and reduction of the heat-affected areas are the most pronounced advantages of the technique compared to effects induced by longer pulses [12], which reflect the merit of the method as a potential tool for fabrication at micro- and nano-scales [13]. These abundant applications require thorough knowledge of the laser interaction with the irradiated material for enhanced controllability of the resulting modification of the target relief.

Femtosecond pulsed laser interaction with matter triggers a variety of timescale-dependent processes, influenced by the fluence and pulse duration [14]; different combinations of those parameters are capable to induce phase transition or material removal. A solid material subjected to ultrashort pulsed laser heating at sufficiently high fluences undergoes a phase transition to a superheated liquid whose temperature reaches $0.90T_{cr}$ ($T_{cr}$ being the thermodynamic critical temperature) [15]. A subsequent bubble nucleation leads to a rapid transition of the superheated liquid to a mixture of vapour and liquid droplets that are ejected from the bulk material (phase explosion). This has been proposed as a material removal mechanism [16, 17]. By contrast, the interpretation of a possible surface modification due to evaporation has been related to the presence of a Knudsen layer adjacent to the liquid-vapor interface and the process has been analysed in numerous works [18-20]. The proposed scenario of modelling material removal is based on a combination of evaporation of material volumes that exceed upon irradiation lattice temperatures close to $0.90T_{cr}$ and evaporation due to dynamics of Knudsen layer, which is a scenario that is yet to be elaborated. Hence, the applied conditions are proposed to be sufficient enough to lead to minimal mass removal while the rest of the material is treated as an incompressible Newtonian fluid. It has to be emphasised that throughout this manuscript the term 'mass removal' is associated with 'mass loss' rather than 'displaced mass' or 'mass ejection'.

A very significant aspect that will be investigated is the development of rippled structures on the semiconductor surface after exposure to repetitive laser pulses. Previous theoretical approaches or experimental observations related to the formation mechanism of these periodic structures were performed in submelting [21] or ablation conditions [10, 22-27]. The proposed theoretical framework will elaborate both optical (the interference of the incident and plasmon waves) and hydrodynamical (capillarity-driven ripple formation) effects. To the best of our knowledge, there is no theoretical interpretation of surface patterning in an intermediate regime, in conditions that lead to a superheated liquid, describe surface modification due to mass redistribution and present surface morphological changes that originate from a phase transition process (resolidification). Although there exists a number of models that investigate short pulsed laser welding processes of metals [18-20], our approach aims to provide a detailed description of the underlying mechanisms of semiconductor surface patterning after irradiation with ultrashort pulses and link ripple formation with development of surface plasmon waves.

The purpose of this article is to introduce a unifying theoretical framework that is capable to describe ultrafast laser induced surface modification and ripple formation by considering that hydrodynamic, and not the so far believed mass removal, effects constitute the main agent that govern the initial morphological changes. Compared to previous established approaches [12, 22, 28, 29], the proposed theory aims to provide an extension based on the spatial dependence (i.e. radial) of the intensity distribution that induces bending of the isothermal lines. The proposed model comprises: i) a heat transfer component that accounts for the particle dynamics and heat conduction phenomena, and (ii) a hydrodynamics component which describes the molten material dynamics and the resolidification process. The laser fluences modelled and experimentally tested were chosen and restricted to lie within the onset of material evaporation, avoiding conditions of mass removal due to ablation. This enabled us to shed light on the key mechanisms governing the embryonic stages of material modification.

## II. THEORETICAL MODEL

### 1. Laser-matter interaction process

Ultrashort-pulsed lasers first excite the charge carriers (electron-hole pairs) in semiconductors while their energy is subsequently transferred to the lattice. The relaxation time approximation to Boltzmann's transport equation [30] is employed to determine the carrier density number, carrier energy and lattice energy. The model assumes an equal number of electron and holes in the solid and no electron photoemission is considered after the laser irradiation [31]. The evolution of the carrier density number $N$, carrier temperature $T_c$ and lattice temperature $T_l$ are derived using the carrier, carrier energy and lattice heat balance equations. Based on this picture the following set of equations determine the temperature and particle dynamics [21, 30, 32, 33]

$$C_c \frac{\partial T_c}{\partial t} = \vec{\nabla} \bullet \left( (k_e + k_h) \vec{\nabla} T_c \right) - \frac{C_c}{\tau_e} (T_c - T_l) + S(\vec{r},t)$$

$$C_l \frac{\partial T_l}{\partial t} = \vec{\nabla} \bullet \left( K_l \vec{\nabla} T_l \right) + \frac{C_c}{\tau_e} (T_c - T_l)$$

$$\frac{\partial N}{\partial t} = \frac{\alpha}{h\nu} \Omega I(\vec{r},t) + \frac{\beta}{2h\nu} \Omega^2 I^2(\vec{r},t) - \gamma N^3 + \theta N - \vec{\nabla} \bullet \vec{J} \quad (1)$$

$$\Omega = \frac{1 - R(T_l)}{\cos \varphi}$$

The laser intensity in Eqs.(1-2) is obtained by considering the propagation loss due to one-, two-photon and free carrier absorption [30]

$$\frac{\partial I(\vec{r},t)}{\partial z} = -(\alpha + \Theta N) I(\vec{r},t) - \beta I^2(\vec{r},t) \quad (2)$$

assuming that the laser beam is Gaussian both temporally and spatially and the transmitted laser intensity at the incident surface is expressed in the following form [21]

$$I(r, z=0, t) = \frac{2\sqrt{\ln 2}}{\sqrt{\pi} \tau_p} E_p e^{-\left(\frac{2r^2}{R_0^2}\right)} e^{-4\ln 2 \left(\frac{t-t_0}{\tau_p}\right)^2} \quad (3)$$

where $E_p$ is the fluence of the laser beam and $\tau_p$ is the pulse duration (i.e. full width at half maximum), $R_0$ is the irradiation spot-radius (distance from the centre at which the intensity drops to $1/e^2$ of the maximum intensity.

In the present work, we assume conditions that lead to temperatures greater than $\sim 0.90 T_{cr}$ (for silicon, $T_{cr} = 5159^0 K$ [34]) for a small part of the material. While this portion is evaporated, a superheated liquid still remains in the system which undergoes a slow cooling. Furthermore, to introduce the phase transition that causes the bulk temperature to exceed the silicon melting temperature $T_m$ ($\sim 1687\ ^0K$), the second equation in Eq.1 has to be modified properly to include the phase change in the sold-liquid interface

$$\left( C_l \pm L_m \delta (T_l - T_m) \right) \frac{\partial T_l}{\partial t} = \vec{\nabla} \bullet \left( K_l \vec{\nabla} T_l \right) + \frac{C_c}{\tau_e} (T_c - T_l) \quad (4)$$

where $L_m$ is the latent heat of fusion. A suitable representation of the $\delta$-function should accomplish (for numerical calculations) a smooth transition between solid and liquid phases [35], therefore the following expression is used

$$\delta (T_l - T_m) = \frac{1}{\sqrt{2\pi} \Delta} e^{-\left[ \frac{(T_l - T_m)^2}{2\Delta^2} \right]} \quad (5)$$

where $\Delta$ is in the range of 10-100$^0$K depending on the temperature gradient. The sign in front of the term that contains the $\delta$-function depends on whether melting or solidification takes place. One aspect that should not be overlooked is that melting of silicon not only induces a solid-to-liquid phase transition, but also alters its properties since the molten material exhibits metal behaviour. Hence, a revised two-temperature model that describes heat transfer from electrons to lattice has to be employed [36] and thereby, for temperatures above $T_m$, Eqs.1 need to be replaced by the following two equations that describe electron-lattice heat transfer

$$C_e \frac{\partial T_e}{\partial t} = \vec{\nabla} \bullet \left( K_e \vec{\nabla} T_e \right) - \frac{C_e}{\tau_E} \left( T_c - T_L \right)$$
$$C_L \frac{\partial T_L}{\partial t} = \frac{C_e}{\tau_E} \left( T_c - T_L \right)$$
(6)

where $C_e$ and $C_L$ are the heat capacity of electrons and lattice (liquid phase), $K_e$ is the thermal conductivity of the electrons, while $\tau_E$ is the energy relaxation time for the liquid phase. The governing equations for the incompressible Newtonian fluid flow and heat transfer in the molten material are defined by the following equations:

(i). for the mass conservation (incompressible fluid):

$$\vec{\nabla} \bullet \vec{u} = 0 \tag{7}$$

(ii). for the energy conservation

$$C_L \left( \frac{\partial T_L}{\partial t} + \vec{\nabla} \bullet \left( \vec{u} T_L \right) \right) = \vec{\nabla} \bullet (K_L \vec{\nabla} T_L) \tag{8}$$

where $K_L$ is the thermal conductivity of the lattice. The presence of a liquid phase and liquid movement requires a modification of the second of Eq.8 to incorporate heat convection. Furthermore, an additional term is presented in the equation to describe a smooth transition from the liquid-to-solid phase (i.e. it will help in the investigation of the resolidification process)

$$C_L \left[ \frac{\partial T_L}{\partial t} + \vec{\nabla} \bullet \left( \vec{u} T_L \right) \right] - L_m \delta \left( T_L - T_m \right) \frac{\partial T_L}{\partial t} = \vec{\nabla} \bullet (K_L \vec{\nabla} T_L) \tag{9}$$

(iii). for the momentum conservation:

$$\rho_L \left( \frac{\partial \vec{u}}{\partial t} + \vec{u} \bullet \vec{\nabla} \vec{u} \right) = \vec{\nabla} \bullet \left( -P\mathbf{1} + \mu \left( \vec{\nabla} \vec{u} \right) + \mu \left( \vec{\nabla} \vec{u} \right)^T \right) \tag{10}$$

where $\vec{u}$ is the velocity of the fluid, $\mu$ is the liquid viscosity, $P$ pressure. $C_L$ and $K_L$ stand for the heat capacity and thermal conductivity of the liquid phase, respectively. It is evident that the transition between a purely solid to a completely liquid phase requires the presence of an intermediate zone that contains material in both phases. In that case, Eq.10 should be modified accordingly to account for a liquid-solid two phase region (i.e. mushy zone) where the total velocity in a position should be expressed as a combination of the fraction of the mixtures in the two phases [37]. Nevertheless, to avoid complexity of the solution of the problem and given the small width of the two phase region with respect to the size of the affected zone a different approach will be pursued where a mushy zone is neglected and transition from to solid-to-liquid is indicated by a smoothened step function of the thermophysical quantities. Furthermore, as it will be explained in the Simulation section, the flow will be assumed to occur for two liquids of significantly different viscosity.

While a complete description would also require the inclusion of buoyancy forces [18-20], its contribution is expected to be minimal and therefore the relevant term will be eliminated. Vapour ejected creates a back (recoil) pressure on the liquid free surface which in turn pushes the melt away in the radial direction. The recoil pressure and the surface temperature are usually related according to the equation [38, 39]

$$P_r = 0.54 P_0 \exp\left( L_v \frac{T_L^S - T_b}{R T_L^S T_b} \right) \tag{11}$$

where $P_0$ is the atmospheric pressure (i.e. equal to $10^5$ Pa), $L_v$ is the latent heat of evaporation of the liquid, $R$ is the universal gas constant, and $T_L^s$ corresponds to the surface temperature. Given the radial dependence of the laser beam, temperature decreases as the distance from the centre of the beam increases; at the same time, the surface tension in pure molten silicon decreases with growing melt temperature (i.e $d\sigma/dT<0$), which causes an additional depression of the surface of the liquid closer to the maximum value of the beam while it rises elsewhere. Hence, spatial surface tension variation induces stresses on the free surface and therefore a capillary fluid convection is produced. Moreover, a precise estimate of the molten material behaviour requires a contribution from the surface tension related pressure, $P_\sigma$, which is influenced by the surface curvature and is expressed as $P_\sigma=K\sigma$, where $K$ is the free surface curvature. The role of the pressure related to surface tension is to drive the displaced molten material towards the centre of the melt and restore the morphology to the original flat surface. Thus, pressure equilibrium on the material surface implies that the pressure in Eq.10 should outweigh the accumulative effect of $P_r + P_\sigma$.

As the material undergoes a solid-to-liquid-to-solid phase transition, it is important to explore the dynamics of the distribution of the depth of the molten material and the subsequent surface profile change when solidification terminates. The generated ripple height is calculated from the Saint-Venant's shallow water equation [40]

$$\frac{\partial H(\vec{r},t)}{\partial t} + \vec{\nabla} \cdot \left( H(\vec{r},t)\vec{u} \right) = 0 \tag{12}$$

where $H(\vec{r},t)$ stands for the melt thickness. Hence, a spatio-temporal distribution of the melt thickness is attainable through the simultaneous solution of Eqs.(1-12).

### 2. Interference of an incident with a surface plasmon wave

Due to an inhomogeneous deposition of the laser energy on the semiconductor as a result of the exposure to a Gaussian-shape beam, the surface of material is not expected to be perfectly smooth after resolidification; further irradiation of the non-planar profile will give rise to a surface scattered wave [25]. According to theoretical predictions and experimental studies, the interference of the incident and the surface wave results in the development of periodic 'ripples' with orientation perpendicular ($p$-polarisation) to the electric field of the laser beam [25-27]. A revised process which that leads to the formation of the surface wave has been also proposed that involves surface plasmons [23, 24] where the ripple periodicity is provided by the expression $\lambda/(\lambda/\lambda_s \pm sin\varphi)$. The involvement of surface plasmon wave related mechanism in the generation of ripples will be employed in this work as the metallic behaviour of silicon at large temperatures allows excitation of surface plasmon waves. The plasmon wavelength, $\lambda_s$, is related to the wavelength of the incident beam through the relations [24]

$$\lambda_s = \lambda \left( \frac{\varepsilon' + \varepsilon_d}{\varepsilon' \varepsilon_d} \right)^{1/2}$$

$$\varepsilon' = \text{Re}\left( 1 + (\varepsilon_g - 1)\left(1 - \frac{N}{n_0}\right) - \frac{N}{N_{cr}} \frac{1}{\left(1 + i\frac{1}{\omega\tau_e}\right)} \right) \tag{14}$$

where $\varepsilon_d$ ($\varepsilon_d=1$) is the dielectric constant of air, $\varepsilon_g$ stands for the dielectric constant of unexcited material ($\varepsilon_g=13.46+i0.048$), $\omega$ is the frequency of the incident beam, $n_o$ is the valence band carrier density ($n_o=5\times10^{22}$cm$^{-3}$), $N_{cr}=m_{eff}\varepsilon_0\omega^2/e^2$ where $m_{eff}$ is the effective electron mass [41].

Although the equations presented in the previous section are still valid, some modification has to be performed to the form of the laser intensity beam due to the interference of the incident and the surface plasmon wave. The final intensity on the surface is provided by the following expression

$$I_{surf}(\vec{r},t) = \left\langle \left|\vec{E}_i + n_{mat}\vec{E}_s\right|^2 \right\rangle e^{-\left(\frac{2r^2}{R_0^2}\right)} e^{-4\ln 2\left(\frac{t-t_0}{\tau_p}\right)^2} \quad (15)$$

where $n_{mat}$ is the refractive index of silicon, $\vec{E}_i = \vec{E}_{i,o}(\vec{r})\exp(-i\omega_i t + i\vec{k}_i \cdot \vec{r})$ and $\vec{E}_s = \vec{E}_{s,o}(\vec{r})\exp(-i\omega_s t + i\vec{k}_s \cdot \vec{r})$ and $\omega_i$, $\omega_s$ are the frequencies of the incident beam (equal to $\omega$) and the surface plasmon wave, respectively. The magnitude of the electric field of the incident wave can be calculated by the expression

$$\frac{E_d}{\tau_p} = \frac{c\varepsilon_0\sqrt{\pi}\left\langle\left|\vec{E}_i\right|^2\right\rangle}{2\sqrt{\ln 2}} \quad (16)$$

while the magnitude of the electric field of the longitudinal surface plasmon wave [42] is taken to be of the order of the electric field of the incident wave. The computation of the time averaged quantity in the final intensity on the surface (Eq.15), yields a contribution proportional to $\cos\left(\sqrt{\vec{k}_i^2 + \vec{k}_s^2 - 2\vec{k}_i \cdot \vec{k}_s}(\sin\varphi)x\right)\vec{E}_{i,o} \cdot \vec{E}_{s,o}$ ($x$-axis is along the direction of the electric field of the incident beam, see Simulation Section), where $\vec{k}_i$ and $\vec{k}_s$ are the wavevectors of the incident and surface plasmon waves, respectively. The above expression indicates that the polarisation of the incident beam is crucial for the determination of the form of the total intensity distribution. Due to the curvature of the initially modified surface after irradiation with one laser pulse, the direction of the surface plasmon waves is geographically correlated. Nevertheless, the energy deposition will be highest in the direction of the laser electrical field and lowest in the perpendicular direction because the electric field of the surface plasmon wave has a larger component along the polarisation of the incident beam. Hence, the energy deposition is strongly correlated to the polarisation of the laser beam and it yields a highest component along the incident electric field polarisation. As a result, for an incident beam with polarisation on the $xz$-plane (Fig.9a), the periodic function produces an optical interference pattern which propagates in parallel to the polarisation vector and it is followed by a spatially and periodically modulated energy deposition. The proposed model suggests that spatial energy deposition will lead to temperature gradients which accounts for the capillarity-driven ripple formation when material is melted and then resolidified.

### III. SIMULATION

The proposed model aims to determine the time-dependent surface profile by solving both momentum and energy equations introduced in the previous paragraphs. It takes into account: (i). a solid-to-liquid phase transition with an energy transfer to the lattice equal to the latent heat (Eqs.1, 6), (ii). the Marangoni effect which describes liquid flow due to temperature gradients [43], (iii). the contribution of recoil pressure and pressure due to surface tension, and (iv) a resolidification process that is quantifiable by computing the liquid-solid interface velocity. Due to a small vertical anticipated surface modification with respect to the size of the laser beam, a finite difference method to solve the Eqs.(1-16) will suffice to produce accurate results. The coordinate system used in the analysis is defined as follows: the $z$-axis is normal to the material surface, the $x$-axis is on the surface with a direction based on that the electric field of the incident beam must reside on the $xz$-plane, and the $y$-axis, again on the material surface (Fig.9a). Due to the axial symmetry (for single pulse irradiation), cylindrical coordinates ($r$ and $z$) can be employed to obtain the carrier and lattice temperatures and the surface modification details. As a result, we can perform a simulation on a rectangular subregion of thickness $W$=5μm and length $L$=30μm is selected. The simulation runs for timepoints in the range from $t$=0 to 15ns in time intervals that vary from 10fs to 50fs to integrate nonequilibrium time history while a substantially larger time step (i.e. 5ps) is followed for post equilibrium evolution. Carriers and lattice temperatures are set to $T=T_0=300^0$K (room temperature) at $t=0$ while the initial concentration of the carrier is set to $N=1\mu m^{-3}$ [30].

The hydrodynamic equations will be solved in the aforementioned subregion that contains either solid or molten material. To include the 'hydrodynamic' effect of the solid domain, material in the solid phase is modelled as an extremely viscous liquid ($\mu_{solid}=10^5 \mu_{liquid}$), which will result into velocity fields that are infinitesimally small. An apparent viscosity is then defined with a smooth switch function similar to Eq.5 to emulate the step of viscosity at the melting temperature. A similar step function is also introduced to allow a smooth transition for the rest of temperature dependent quantities (i.e. heat conductivity, heat capacity, density, etc) between the solid and liquid phases. For time-dependent flows, a common technique to solve the Navier-Stokes equations is the projection method and the velocity and pressure fields are calculated on a staggered grid (Fig. 1a) using fully implicit formulations [44, 45]. The inset in Fig.(1a) illustrates the different locations at which the velocity ($v,u$) and pressure $P$ fields are calculated at a timepoint $t=t^l$. More specifically, the horizontal and vertical velocities are defined in the centres of the horizontal and vertical cells faces, respectively where the pressure and temperature fields are defined in the cell centres. Similarly, all temperature dependent quantities (i.e. viscosity, heat capacity, density, etc) are defined in the cell centres. An explicit solution method is employed to solve Eq.10 and compute temperature values at subsequent time points. The size of the horizontal side of the computation grid is 0.1nm while the vertical side is taken to be equal to 0.01nm.

During the ultrashort period of laser heating, heat loss from the upper surface of target is assumed to be negligible. As a result, the heat flux boundary conditions for carriers and lattice are zero throughout the simulation while a zero flux at $r=0$ must be also imposed. Furthermore, it is assumed that only the top surface is subjected to the Gaussian-shape laser beam of an irradiation spot-radius $R_0=15\mu m$. Regarding the momentum conditions at the boundaries, we impose the following constraints:

1. $u_r=0$, on the symmetry axis $r=0$,
2. $\vec{u}=0$, on the solid-liquid interface (non-slipping conditions),
3. $\mu \frac{\partial u_r}{\partial z} = \frac{\partial \sigma}{\partial T_s} \frac{\partial T_s}{\partial r}$, on the upper flat free surface ($T_s$ is the surface temperature) [46], and

   $\mu \frac{\partial u_\tau}{\partial n} = \frac{\partial \sigma}{\partial T_s} \frac{\partial T_s}{\partial \tau}$, on the rest curved free surface [39], where $\tau$ and $n$ are the surface tangent direction and normal component, respectively.

The aforementioned boundary conditions are valid when the first incident pulse irradiates the initial planar surface. The first boundary condition excludes transverse component of the velocity on the symmetry axis ($r=0$). The last expression describes the shear stress which is due to the surface tension gradient and it is exerted on the free surface and it is important to incorporate the effect of Marangoni flow. The temperature-dependent parameters that are used in the numerical solution of the governing equations are listed in Table I. In order to increase accuracy, the model geometry is divided into two regions: a subdomain in which solid phase dominates and velocity fields are minimal and another subdomain where both momentum and energy equations are solved with an increased tolerance level.

For irradiation with subsequent pulses, we note that the incident beam is not always perpendicular to the modified profile, therefore the surface geometry influences the spatial distribution of the deposited laser energy. Hence, the laser irradiation reflected from the profile slopes can lead to light entrapment among the formed structures where the laser fluence is modified. Typical Fresnel equations are used to describe the reflection and transmission of the incident light. Due to multiple reflection, absorption of the laser beam is modified [47] and thereby a ray tracing method is employed to compute the absorbed power density while a similar methodology is ensued to estimate the proportion of the refraction by applying Snell's law. With respect to the numerical scheme used to simulate dynamics of velocity and pressure fields and all thermophysical quantities in Eqs.(1-12), a similar procedure is ensued in the event of subsequent pulses, however in this case the interaction with the modified surface profile induced by the first pulse due to the hydrodynamic motion of the molten material and its subsequent resolidification, should be taken into account. While second order finite difference schemes appear to be accurate for $NP=1$ where the surface profile has not been modified substantially, finer meshes and higher order methodologies are performed for more complex and profiles [48, 49]. The calculation of the pressure associated to the surface tension requires the computation of the temporal evolution of the principal radii of surface curvature $R_1$ and $R_2$ that correspond to the convex and concave contribution, respectively [50]. Hence the total curvature is computed from the expression $K=(1/R_1 +1/R_2)$. A positive radius of the melt surface curvature corresponds to the scenario where the centre of the curvature is on the side of the melt relative to the melt surface.

Regarding the material removal simulation, in each time step, lattice and carrier temperatures are computed and if lattice temperature reaches ~$0.90T_{cr}$, mass removal through evaporation is assumed. In that case, the associated nodes on the mesh are eliminated and new boundary conditions of the aforementioned form on the new surface are enforced. In order to preserve the smoothness of the surface that has been removed and allow an accurate and non-fluctuating value of the computed curvature and surface tension pressure, a fitting methodology is pursued. Fig.1b illustrates simulating data and the resulting curve after data fitting to produce a smoothed curve. To facilitate assessment of data fitting accuracy, the resulting smoothing is sketched in an enlarged rectangular region of Fig.1b (Fig.1c).

## IV. EXPERIMENTAL DETAILS

Experiments were performed with a femtosecond Ti:Sapphire laser system operating at a wavelength of 800 nm and repetition rate of 1 kHz. The pulse duration was set to 430 fs and measured by means of cross correlation techniques. A pockels cell controlled the repetition rate and the number of the pulses that irradiated the silicon surface. The beam was perpendicular to the silicon substrate giving a spot diameter of 30µm located inside a vacuum chamber evacuated down to a residual pressure of $10^{-2}$ mBar. The laser fluence calculated from the beam waist ($1/e^2$) was 0.37 J/cm$^2$, chosen to assure a minimal mass removal for a single pulse irradiation. This is performed by means of atomic force (AFM- Nanonics Multiview-4000) and Field emission scanning electron microscopy FESEM – JEOL JSM-7000F (images of the profiles of the various modification spots obtained.).

## V. RESULTS AND DISCUSSION

The theoretical model is firstly examined to derive the time evolution of carrier (electron-hole for semiconductors and electrons for metals) and lattice temperatures for one laser shot ($NP$=1) of fluence $E_p$=0.37J/cm$^2$ and pulse duration $\tau_p$=430fs. The fluence value used in the simulation is appropriate for investigation of thermophysical effects when minimal mass loss removal is assumed and the selection was based on the requirement of exploration of morphological changes in sub-ablation conditions. Fig.2a depicts the evolution of the temperatures that occur at $r$=0, $z$=8.2nm, which corresponds to lattice points that have reached temperatures higher than $0.90T_{cr}$. Fig.2b illustrates the lattice temperature spatial distribution at $t$=5ps, which indicates that the material is divided in three regions: the first region (inside the region defined by the boundary in *white*) corresponds to material that has been evaporated due to the fact that it reaches temperatures above that of $0.90T_{cr}$; the second one refers to a superheated liquid with temperatures in the range [$T_m$, $0.90T_{cr}$]; the third region corresponds to material in solid phase with temperatures less than $T_m$. The choice of $t$=5ps was based on the fact that it corresponds to the maximum calculated material removal (as the attained temperature reaches a maximum value) for the aforementioned laser beam characteristics and the radial size of the approximate removed portion is equal to 6.5µm. According to Fig.2b, the theoretically calculated depth of the evaporated material is approximately equal to 8.2nm.

To estimate surface modification due to the cooling mechanism of the superheated liquid, a detailed investigation of the liquid-to-solid phase transition is required. It is noted that other works focused on the consideration that the onset of melting involves homogeneous or heterogeneous nucleation of the molten phase at lattice defects [51, 52]. According to these works, atomistic simulations predict an increased melting front speed dependent on the degree of overheating which potentially has an impact on the simulation results. It is evident that the incorporation of a model with molecular dynamics simulations and lattice instabilities potentially results in a complementary description of the general case of material especially in cases where material has structural instabilities. Nevertheless, the aim of the present investigation is aimed to characterize a simpler scenario in which melting is basically associated with a homogeneous nucleation once melting temperature has been reached and compare theoretical results to experimental observations.

Melting of material entails elaboration of the solidification process and therefore hydrodynamics effects influenced by recoil pressure, surface tension pressure and surface tension gradient have been investigated. To examine the interplay of the various types of pressure and their resulting effect on the surface modification, the evolution of the melt recoil pressure and surface tension pressure and their relation are assessed. To determine which is the dominant mechanism of surface modification and the hill formation process, the contribution of each term is required to be estimated and it known that the speed of the melt flow is related to the recoil pressure through the expression $P_r = 1/2 \rho_m (u_m)^2$ (i.e. the subscript *m* characterizes the molten material) [18]. Furthermore, the exponential dependence of the recoil pressure on surface temperature causes a large radial gradient for the recoil pressure and this gradient accounts for the melt flow in the radial direction. Moreover, the outward melt leads to the creation of a small protrusion at the edge of the affected region that undergoes melting whose curvature varies substantially with respect to the curvature of the surface closer to the beam centre. Fig.2c shows the temporal evolution of all types of pressure at the edge of the zone that undergoes melting. The illustration indicates that the recoil pressure effect is larger than the pressure due to surface tension in the early stages of resolidification that produces a small melt ejection towards the edge of the zone. By contrast, at later stages and before the end of the resolidification process the recoil pressure is outweighed by the increase of the pressure due to surface tension. Hence, the resulting pressure will produce a protrusion of size that increases in a way that forces due to recoil pressure and surface tension will balance one another. Therefore, material ejection out of the irradiated zone will not further occur till the material cools to temperatures below the melting point.

Figs. (3a)-(3c) illustrate the evolution and the transient behaviour of the movement of the molten volume of the material while the thick arrows show the direction of the associated velocity fields at *t*=0.1,1, and 10ns, respectively. Recoil pressure has initially an important impact in the surface depression. Figs (3a)-(3c) show that molten material displacement due to the combination of recoil pressure and surface tension gradient results in a decrease of the melt thickness and melting of a new portion of the material. Temperature gradient coupled with surface tension gradient induces an outward and upward flow (i.e Marangoni flow). Furthermore, recoil pressure makes the molten material flow upwards. Simulation results show that upward melt flow delivers both heat and molten material towards the edge of the depressed zone. The displaced material will produce a convex profile that will undergo two oppositely directed pressures (one due to the surface tension and another form decreasing recoil pressure) that will collide and a protrusion will be formed at the edge of the affected region. A liquid temperature decrease at larger distances from the laser beam centre leads to increasing liquid surface tension which is an additional reason why liquid is pulled away from the centre. Furthermore, a surface tension variance leads to the development of clockwise flow and capillary waves which eventually results into surface depression in regions of higher temperatures [40, 53]. Figs (3d)-(3f) illustrate the temperature distribution at different time-points which is associated to the heat flow inside the molten material.

The employment of the Navier-Stokes equation and estimation of the liquid-solid interfacial location leads to the surface profile illustrated in Fig.4 which shows that there exists an increased surface depression with respect to the initially flat profile. In comparison to the surface profile resulted from mass evaporation, phase transition of the molten material induces an enhanced radially dependent depression and for mass conservation reasons a hill protrusion is pronounced above the surface. Fig.4 illustrates the theoretically predicted crater formation after a single pulse irradiation and the maximum depth of the crater is estimated to be equal to about 20nm. Due to axial symmetry a similar process and a symmetric profile is expected for the other half of the material (results not shown). Both experimental observations and the theoretically predicted value (i.e. 8.2nm) are very low compared to the laser beam penetration depth which ranges from 2.6μm to 4.4μm for the range of lattice temperatures during the laser pulse illumination (i.e. penetration depth is estimated from the inverse interband absorption coefficient and it decreases monotonically with increasing temperature). Furthermore, a top view of the SEM image of the spot attained after irradiation with one pulse is illustrated in Fig.5a where enhancement of the crater formation is performed by reversing the contrast of the intensity image (decreasing depth is represented by darker pixels). A cross-line (*dashed* line in Fig.5a) exhibits a roughened and steep crater with a precise mass removal and a small pronounced hill protrusion (Fig.5b). Although similar observations including ablated material and a protruding hill in the periphery of the affected region (attributed to recoil pressure of the ablated material) have been observed in ablation experiments [22, 28], the proposed model predicts similar results in sub-ablation conditions. A significant conclusion inferred from Fig.5b is that there is a minimal (but existing) mass removal from the material as a result of the irradiation. To avoid misinterpretation or mass removal underestimation and exclude any preferential orientation of mass displacement, image analysis was performed in the whole region that depicts surface modification in Fig.5a (and similar images produced from a single irradiation) and similar profiles (Fig.5b) were attained. Hence, quantification of the image provides a justification that a simple mass redistribution is not capable to explain the process. The depth of the evaporated material computed from the simulation (i.e. equal to 8.2nm), is comparable to the average depth measured in the AFM image of the single-pulse spot (i.e. approximately equal to 5.4nm). More specifically, Figs.(5c,d) illustrate a portion of the modified profile and a cross section, respectively.

Irradiation of the non-flat surface with a second pulse (which is performed after completion of surface solidification due to the first pulse) give rises to a plasmon wave excitation which interferes with the incident field and produces a spatially modulated energy deposition as explained in Sec.II.2. A non uniform temperature distribution along the modified surface produces again locally surface tension gradients that in combination of a balance between recoil pressure and surface tension pressure are responsible for the rise and depression of molten surface. However, interference of the incident beam with surface plasmon waves leads to a periodic melting of the initially formed profile. The transient behaviour of the molten portion of the material is characterised by a clockwise flow of molten material and periodic structures are developed with an orientation perpendicular to the electric filed. The interference of the initial beam with the surface plasmon wave will destroy the axial symmetry of the system and thereby surface morphology will not exhibit a cylindrical symmetry. As a result, for *NP*=2, a three dimensional solution of Eqs.(1-16) was employed as explained in the Simulation section. Calculation of the absorbed energy due to multiple reflection and application of Fresnel expressions shows an enhanced energy density locally. In order to interpret the morphological changes and associate them to thermophysical processes, the effect of the nonuniform energy density deposition has to be investigated. It is evident that a higher recoil pressure is exerted on a modified surface profile which results in an increased flow velocity and force applied to the molten material. On the other hand, due to a periodic deposition of the laser density, a recoil pressure will also be a periodic function and spatially modulated. Moreover, the values of local maxima decrease towards the edge of the affected region which indicates that at positions where the recoil pressure function has local maxima, upward flow will be stronger on the side which is closer to the centre. By contrast, close to the points which are related to local minimum for the recoil pressure, the profile is characterised by an increased curvature that results to higher pressure due to surface tension. Hence, the collision of the two opposing types of pressures yields pronounced protrusions.

Figs. 6(a)-(d) illustrate the transient behaviour of the movement of the molten volume and the distribution of the temperature fields after a second pulse irradiates the material (*NP*=2) at *t*=1, and 10ns, respectively. To elucidate the behaviour and facilitate the process observation, figures are restricted only to a small portion of the affected zone (equal to a 2.1μm size along the *x*-axis). The thick arrows in Figs. 6(c,d) show the direction of the associated velocity fields. The flow of the molten material in Figs 6(c,d) demonstrates that the aforementioned interpretation of ripple wavelength with respect to the number of pulses is more evident during the formation of the ripple profile. Figs 6(e,f) illustrate the periodic behaviour of the spatial dependence of the recoil pressure at two different timepoints, *t*=1 and 10 ns, respectively. According to the illustrations, at the bottom of each well, recoil pressure exhibits local maxima and it is stronger on the left side of the well.

To investigate incubation effects and the evolution of the ripple periodicity, the surface profile modification is probed after irradiating the material with more pulses. Fig.7a illustrates the ripple formation for 2, 4, and 8 pulses on the *xz*-plane at *y*=0 and it is evident that the spot size, depth and amplitude increase monotonically with number of pulses. We notice that for *NP*=4, ripples develop above the initial level of the surface profile (i.e flat surface) while for smaller values of *NP*, ripples form below *z*=0. An increase of the spot depth with an increase in the number of pulses is explained by the evaporation of the top layer of the surface profile after irradiation with subsequent pulses (i.e. every new pulse irradiates a modified surface profile and it will cause a further mass removal). Furthermore, a repetitive exposure to pulses leads to ripples with larger amplitude due to an increased gradient of energy deposition as number of pulses grows which causes the ripples to be more pronounced. Ripple height is spatially dependent due to the periodic variation of the gradient of the surface tension and temperature gradient that generates the creation of the periodic structures. Bigger variations are expected closer to the centre of the laser beam yielding structures with bigger peaks.

The average ripple period after laser irradiation with four pulses is estimated to be equal to 736nm. Fig.7b illustrates the theoretically predicted three dimensional spatial dependence of the surface patterning for *NP*=8 in one quadrant where there is a pronounced rippled surface. Ripple horizontal profile resembles the result illustrated in Fig.7a. A cross section (*dotted* line in Fig.7b) along a ripple indicates a spatial decrease followed by an increasing behaviour while along the well (*dashed* line in Fig.7b) it increases monotonically before the occurrence of a small protrusion at the edge of the affected region. The two distinct types of behaviour are exhibited in Fig.7c and Fig.7d, respectively where fluctuations have been removed by filtering simulation results using a mean filter.

In order to measure and investigate quantitatively the modifications in the surface morphology, theoretical simulations are tested against the experimental data. Fig.8a illustrates a top-view SEM image of the spot attained after irradiation with four pulses and the intensity profile along a perpendicular to ripples direction (*dashed* line) was used to estimate ripple periodicity. Interestingly, due to spot or laser beam irregularities, periodicity is not obvious and thereby attention has been restricted to areas where some conclusive measurements are obtainable. Therefore the selected region (*dashed* crossline in Fig.8a) based on that it is characterised by small intensity variances and a well-defined ripple profile can be attained. It is evident that for a *p*-polarised beam the ripples develop perpendicularly to the electric field of the incident beam which also has been reported in laser induced welding, hardening and annealing [22, 25, 27]. Image analysis techniques including a fast Fourier transform (FFT) low pass filter were performed to remove noise and produce the intensity profile without significant fluctuation. Estimation of ripple periodicity is achievable through the computation of the horizontal distance between the pronounced local minima (Fig.8b) which yields an average experimental value of 738±6nm. The analysis of the SEM image along the cross-line exhibits a morphology with a crater in the centre of the heat affected region, a small protrusion in the periphery and an interference pattern which is similar to the theoretically predicted profile (Fig.7a). It should be noted, though, that this intensity-based profile provides accurate information only for the dimensions on the sample plane, while in the vertical direction it gives relative rather than the absolute dimensional values. A depth-related analysis approach through a processing of the AFM image and plot of the radial dependence of the depth across selected cross sections (*dashed* line in Fig.8c) yields a similar value for the period. Although the ripple period is more accurate by analysing the AFM-obtained surface features, the image acquisition procedure performed by the pointed tip produces estimation errors in computing the correct depth and thereby the peak of the ripples is used to compute periodicity. Fig.8d illustrates the ripple amplitude along a cross section (*dashed* line in Fig.8c) which shows a surface patterning with protruded ripples above the unaffected flat surface of the material which conforms to the theoretical results for *NP*=8 rather than *NP*=4 (Fig.7a). The difference can be attributed to the discrepancy between the experimental and theoretical laser beam fluence values. It is evident that an increased energy deposition induces larger temperature gradients and ripple amplitudes which means that a possibly larger experimental fluence value can account for the observed profile at a lower number of pulses.

Figs.9a,b illustrate a correlation between the surface plasmon wavelength, ripple wavelength, carrier density and number of pulses. It is evident (Eq.14) that an increase of the carrier density causes an increase in the surface plasmon wavelength (Fig.9a). By contrast, as the number of pulses increase, the absorbed energy due to multiple reflection increases locally (in the neighborhood of the wells), however, due to a larger surface roughness in other parts of the surface energy deposition decreases which influences the material dielectric constant. The resulting decrease of the absorbed energy will lead to a lower carrier density distribution and lattice temperatures. Furthermore, with respect to the molten material dynamics, the contribution of the recoil pressure in conjunction with surface tension pressure will push the fluid back towards the centre of the irradiated zone which leads to an enhanced decrease of the ripple wavelength compared to the surface plasmon wavelength (Fig.9b). More specifically, it is evident that as the profile becomes deeper and steeper (with increasing number of pulses), pressure due to surface tension near the protrusions play an increasingly predominant role and the displacement of molten material towards to the positions of local maxima of the recoil pressure is expected to be more pronounced. Hence, irradiation of the material with a higher number of pulses lowers the ripple wavelength as protruded hills are pushed towards the wells. As a result, an interpretation of the correlation between the ripple wavelength and the number of pulses is proposed which is based completely on hydrodynamics grounds. Despite an apparent small discrepancy between the ripple and the surface plasmon wavelengths, the change has to be emphasised and flow dynamics appear to explain the trend. Moreover, the resulting ripple wavelength and the proximity of its value to the surface plasmon wavelength indicate that although the surface structure periodicity is determined

predominantly by optical effects (i.e. interference), our simulations signify that hydrodynamics and the evolution of the melt flow are capable to maintain surface structures (subwavelength ripples). Hence, the influence of the recoil pressure, surface tension pressure, capillary effects and the subsequent solidification fail to lead to the destruction of a patterned surface with an optically prescribed wavelength.

The monotonic decrease of the ripple subwavelength periodicity with increasing pulse number (Fig.9b) has also been reported in other works that assume ablation [23], however, variance is minimal around the estimated value for *NP*<8. Furthermore, Fig.9b illustrates a comparison between the theoretically computed ripple period and the experimentally observed values for *NP*≤8. The sketch indicates that the theoretical results are within the experimental error and therefore the model offers an adequate description of the underlying mechanism. Certainly, the discrepancy of the results can be partly attributed to the difficulties to create in the laboratory the exact conditions of the simulations (i.e. eradication of experimental errors). Nevertheless, the theoretical framework reveals with a good accuracy the tendency of the expected results.

The results presented in this work suggest that our model provides a very good insight into the process that dictates the formation of ripples and yields significant quantitative details of the underlying mechanism. Our approach provides a detailed characterisation of the morphology change, ripple characteristics, spot size and propose the hydrodynamic factor to be of predominant importance. In comparison with alternative approaches regarding the interpretation of the ripple formation [24, 25, 54], our theoretical framework aims to provide a complementary methodology by incorporating a variety of ultrashort-pulsed induced effects. It is capable to incorporate a surface plasmon wave-induced ripple formation mechanism by considering a plasmon wave interference with the incident beam, the spatial modulation of the deposited energy on the material surface and the influence of the hydrodynamic factor. Furthermore, a revised version of the theoretical model that incorporates ablation and hot electron explosion effects potentially could provide a smooth transition to understand the underlying mechanism of a yet unjustified groove formation [22] or spike development [55].

## VI. CONCLUSIONS

A detailed theoretical unified model was presented which is aimed to account for the surface modification and the plasmon-generated-periodic surface structure formation (ripples) observed during semiconductor irradiation with ultrashort laser pulses in sub-ablation conditions. It contains heat transfer and hydrodynamics components that describe the particle dynamics, carrier excitation conduction phenomena and the morphology modification, respectively. By choosing beam characteristics that result in minimal mass loss, we demonstrated that it is likely to modulate the surface profile upon laser irradiation. Theoretical simulations and experimental observations indicate that temperature dependent surface tension gradients and recoil pressure are capable to generate molten material movements that lead to the observed morphology changes after capillary wave freezing. Elucidation of the underlying mechanism that dictates surface modification in semiconductors will allow control and alteration of their optoelectronic properties. A similar approach can be ensued to investigate related phenomena in other types of materials such as metals and dielectrics which enhance the applicability of the model.

**Acknowledgement**

This work was supported by the Integrated Initiative of European Laser Research Infrastructures LASERLAB-II (Grant Agreement No. 228334). The authors acknowledge Dr. E. Spanakis and Ms. Aleka Manousaki for their support with the Atomic Force and Scanning Electron Microscopes, respectively.

TABLE I. Model parameters for Si.

| Solid Phase | | |
|---|---|---|
| **Quantity** | **Symbol (Units)** | **Value** |
| Initial temperature | $T_0$ (°K) | 300 |
| Electron-hole pair heat capacity | $C_c$ (J/μm³ K) | $3Nk_B$ |
| Electron-hole pair conductivity [56] | $K_c$ (W/μm K) | $10^{-6} \times (-0.5552 + 7.1 \times 10^{-3} \times T_c)$ |
| Lattice heat capacity [57] | $C_l$ (J/μm³ K) | $10^{-12} \times (1.978 + 3.54 \times 10^{-4} \times T_l - 3.68\, T_l^{-2})$ |
| Lattice heat conductivity [57] | $K_l$ (W/μm K) | $0.1585\, T_l^{-1.23}$ |
| Band gap energy [58] | $E_g$ (J) | $1.6 \times 10^{-19} \times (1.167 - 0.0258\, T_l/T_0 - 0.0198\, (T_l/T_0)^2)$ |
| Interband absorption (800nm) [58] | $\alpha$ (μm⁻¹) | $0.112\, e^{T_l/430}$ |
| Two-photon absorption (800nm) [58] | $\beta$ (sec μm/J) | $9 \times 10^{-5}$ |
| Reflectivity (800 nm) [59] | $R$ | $0.329 + 5 \times 10^{-5}\, T_l$ |
| Auger recombination coefficient [30] | $\gamma$ (μm⁶/sec) | $3.8 \times 10^{-7}$ |
| Impact ionisation coefficient [30] | $\theta$ (sec⁻¹) | $3.6 \times 10^{10}\, e^{-1.5 E_g/k_B T_c}$ |
| Free carrier absorption cross section (800nm) [58] | $\Theta$ (μm²) | $2.9 \times 10^{-10}\, T_l/T_0$ |
| Energy relaxation time [56] | $\tau_e$ (sec) | $\tau_{e0}\left[1 + \left(\dfrac{N}{N_{cr}}\right)^2\right]$, $\tau_{e0}=0.5$ps, $N_{cr}=2\times10^9$ μm⁻³ |
| **Molten Phase** | | |
| Electron heat capacity [60] | $C_e$ (J/μm³ K) | $10^{-16} \times T_e$ |
| Electron conductivity [60] | $K_e$ (W/μm K) | $67 \times 10^{-6}$ |
| Lattice heat capacity [61] | $C_L$ (J/μm³ K) | $1.06 \times 10^{-12}\, \rho_L$ |
| Lattice heat conductivity [62] | $K_l$ (W/μm K) | $(0.5 + 2.9 \times 10^{-4}(T - T_m)) \times 10^{-4}$ |
| Density [63] | $\rho_L$ (gr/μm³) | $10^{-12} \times (3.005 - 2.629 \times 10^{-4}\, T_L)$ |
| Dynamic viscosity [61] | $\mu$ (gr/μm sec) | $3.53 \times 10^5\, \rho_L$ |
| Surface tension [62,64] | $\sigma$ (N/m) | (1) $0.7835 - 0.65 \times 10^{-3}(T - T_m)$, for T<1773K<br>(2) $-1.94789 \times 10^{-11}(T - 5108.13)^3 + 0.00238748$, for T>1773K and T<$0.9\, T_{cr}$<br>(3) $3.70923 \times 10^{-7}(T_{cr} - T)^{1.5}$, for T>$0.9\, T_{cr}$ |
| Energy relaxation time [30] | $\tau_E$ (sec) | $10^{-12}$ |
| Melting temperature [65] | $T_m$ (K) | 1687 |
| Boiling temperature | $T_b$ (K) | 3514 |
| Critical point temperature | $T_{cr}$ (K) | 5159 |
| Latent heat of melting [65] | $L_m$ (J/μm³) | $4206 \times 10^{-12}$ |
| Latent heat of evaporation [60] | $L_v$ (J/μm³) | $32020 \times 10^{-12}$ |

Fig 1 (Color online): (a) Staggered grid: pressure-nodes are shown as black dots, while *u*-component and *v*-component of the velocity are computed at the centre of the horizontal and vertical solid lines, respectively, (b) Smoothing of simulating data after a single illumination, (c) An inset of the enlarged area (rectangular region in (b)).

Fig 2 (Color online): (a) Time evolution of carrier and lattice temperatures at $r=0$ and $z=8.2$nm, (b) Spatial distribution of lattice temperatures at $t=5$ps (*white* line represents boundary of evaporated region), (c) temporal evolution of recoil ($P_r$) and surface tension ($P_\sigma$) pressures.

Fig 3: Surface profile and flow pattern (a-c) and temperature distribution (d-f) at $t=0.1, 1, 10$ns after first irradiation (theoretical results).

Fig 4 (Color online): Theoretical results for *NP*=1: Surface modification as a function of radial distance (axial symmetry).

Fig 5: Experimental results for *NP*=1: (a) SEM intensity image, (b) Intensity profile and spatial dependence along *dashed* cross-line shown in (a), (c) AFM image (d) a cross section of the AFM image.

Fig 6: Simulations for temperature distribution (a,b) and flow pattern (c,d) at $t=1, 10$ns after irradiation with two pulses (*NP*=2). Radial dependence of recoil pressure at $t=1$ns (e) and $t=10$ns (f), respectively.

Fig 7 (Color online): Theoretical results for: (a) Ripple formation after repetitive irradiation (*NP*=2, *solid* line) with *NP*=4 (*dashdot* line), *NP*=8 (*dashed* line) at *xz*-plane (*y*=0), (b) Ripple formation (simulation results) for *NP*=8, (c) Spatial dependence of depth along ripple (*dotted* line in (b)), (d) Spatial dependence of depth along well (*dashed* line in (b)).

Fig 8 (Color online): Experimental results for *NP*=4: (a) SEM intensity image (shows the polarization of the incident beam-on the *x-z* plane and parallel to *x*), (b) Intensity profile spatial dependence along *dashed* cross-line shown in (b), (c) AFM intensity image, (d) Intensity profile spatial dependence along *dashed* cross-line shown in (c).

Fig 9: (a) Theoretical results for surface plasmon wavelength as a function of carrier density, (b) Theoretical results and experimental observations of the ripple wavelength's dependence on number of pulses.

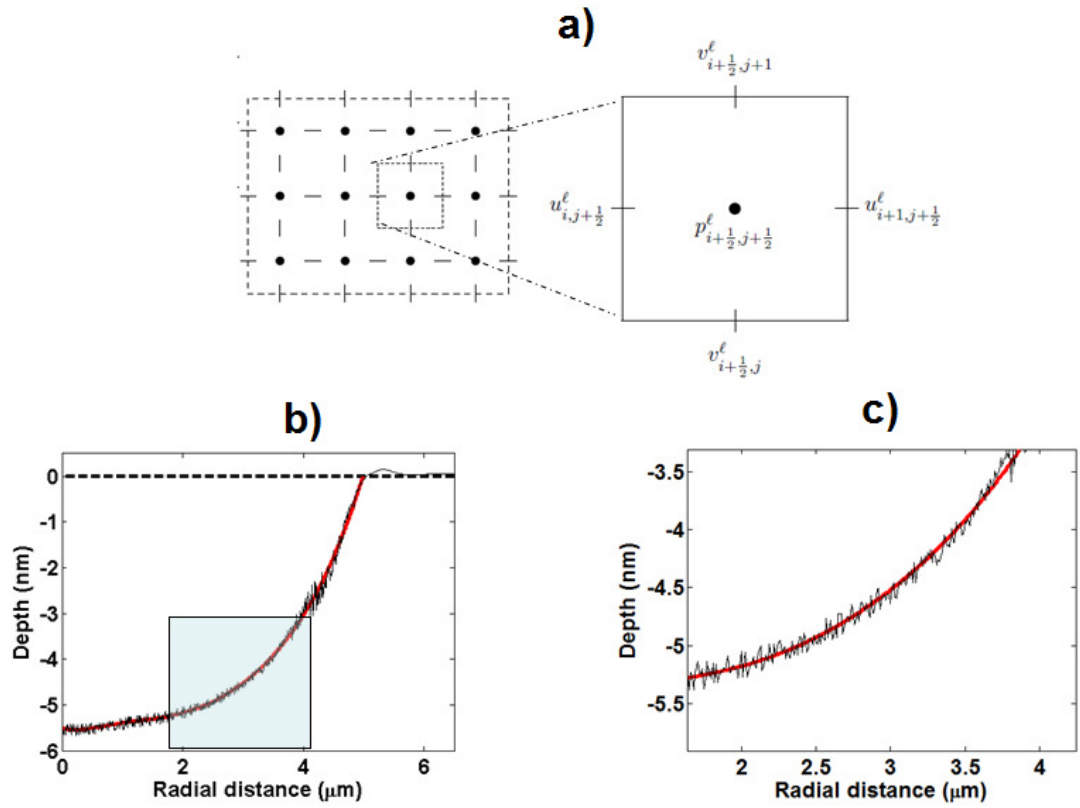

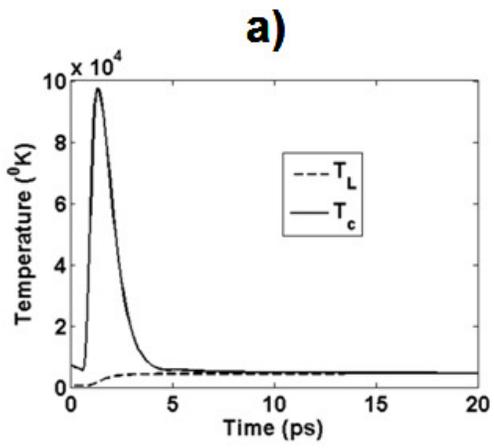
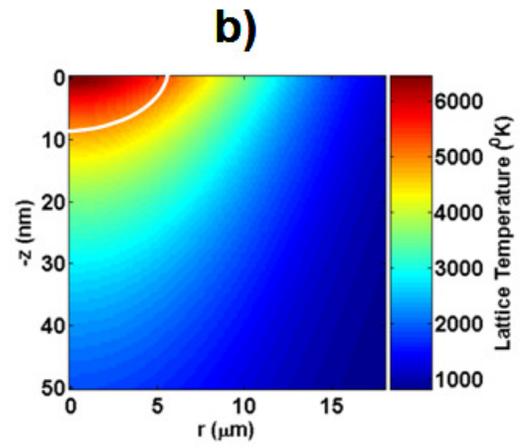
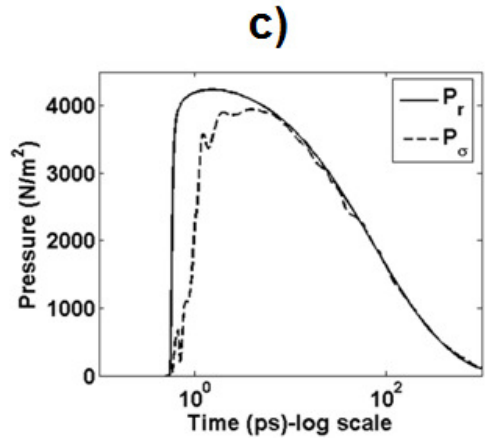

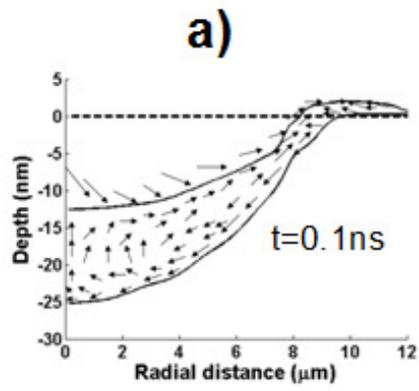
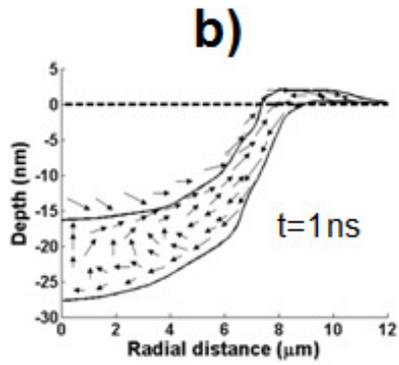
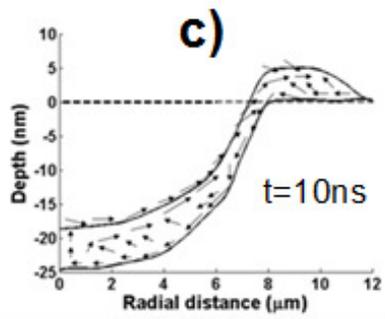
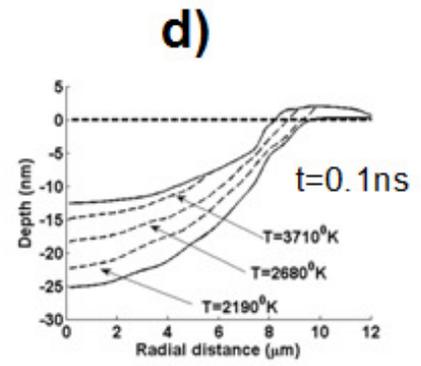
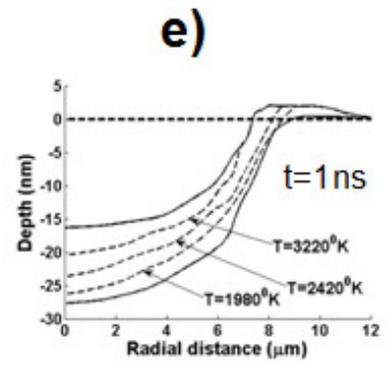
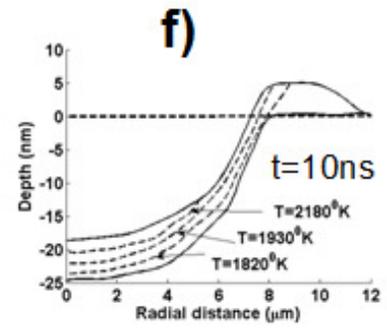

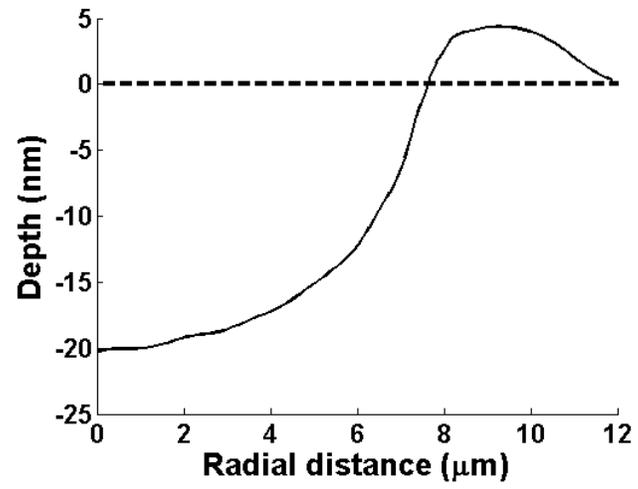

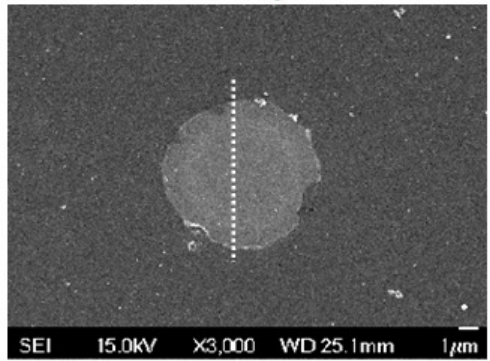 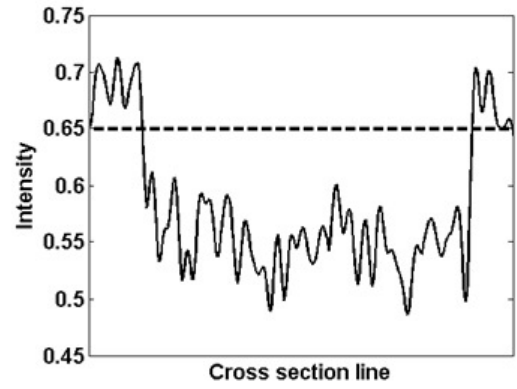
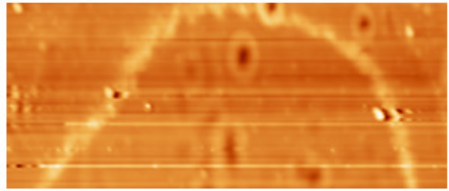 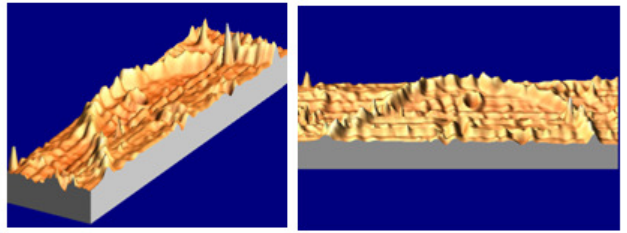

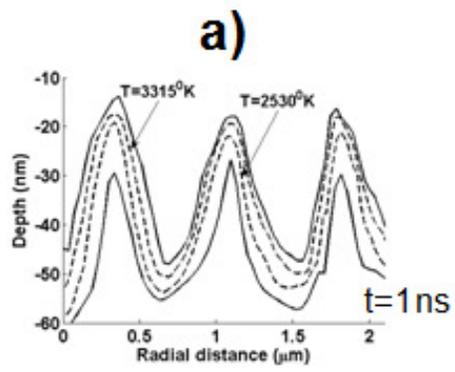
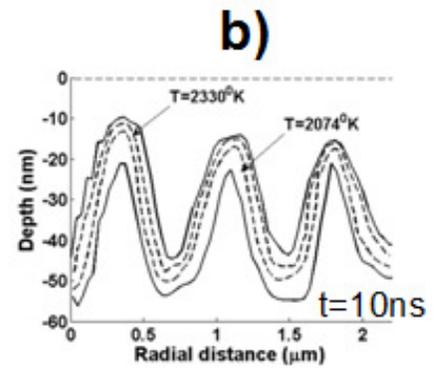
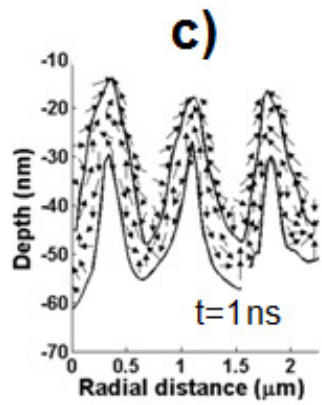
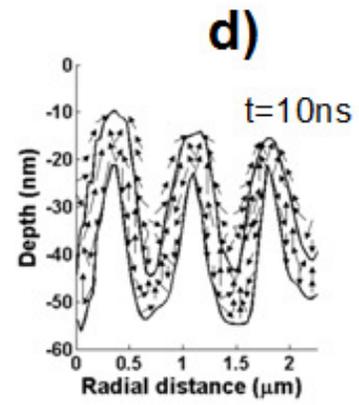
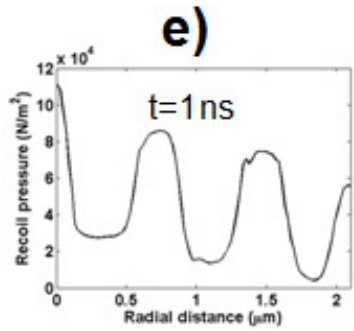
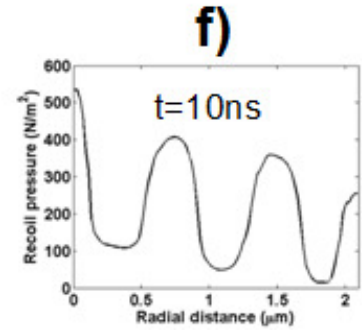

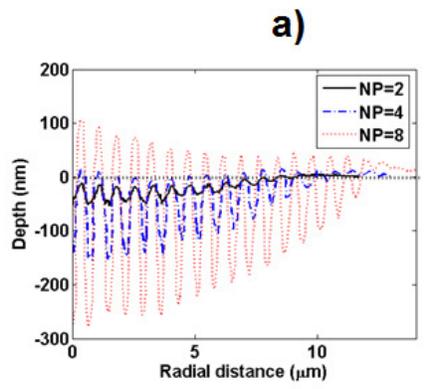 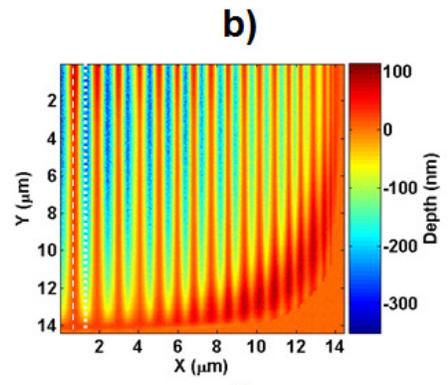
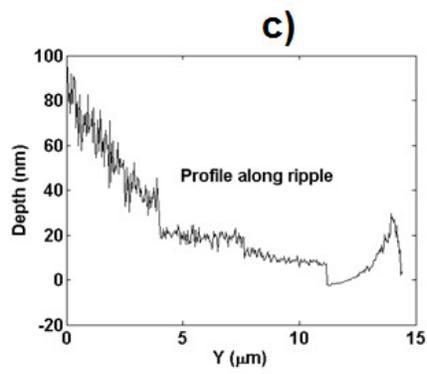 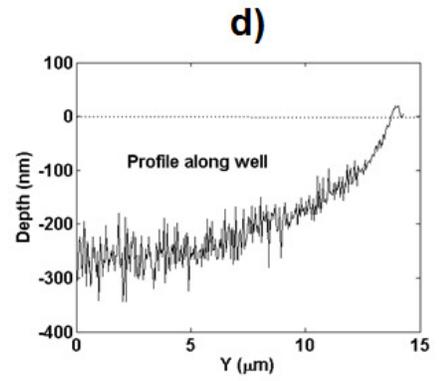

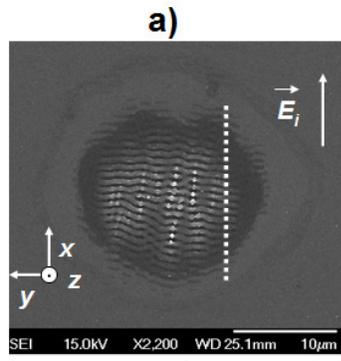 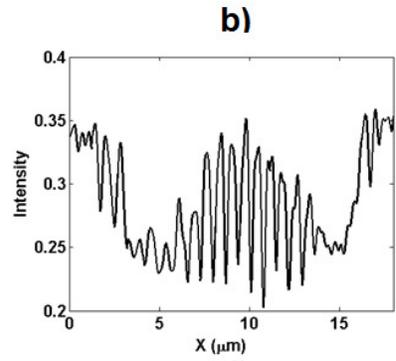
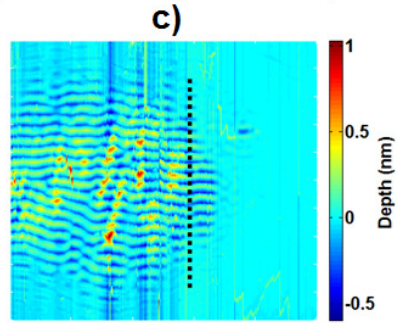 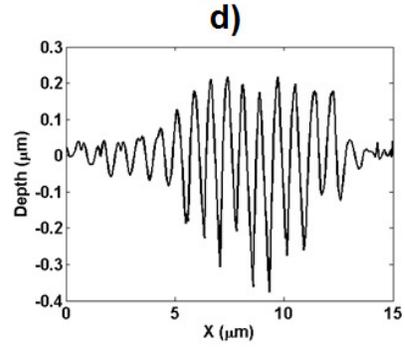

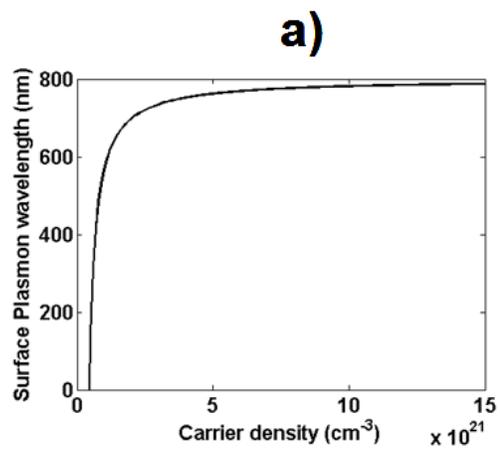 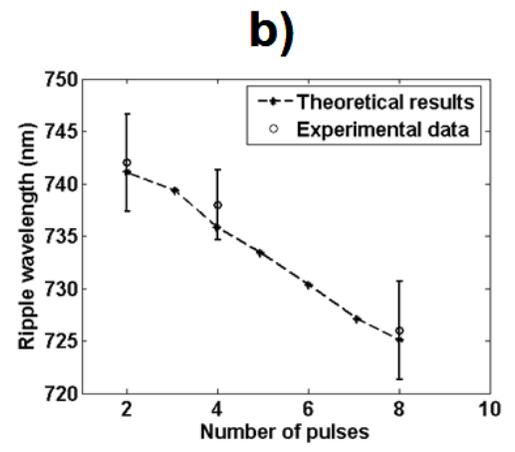

♣ Corresponding author, E-mail address: tsibidis@iesl.forth.gr